\documentclass[doublecol]{epl2}
\usepackage{amsmath}
\usepackage{color}
\usepackage{graphicx}
\usepackage{subfigure}
\usepackage{stfloats}
\usepackage{lipsum} 

\title{$\mathbf{k}$-core percolation on interdependent and interconnected multiplex networks }
\shorttitle{$\mathbf{k}$-core percolation on interdependent and interconnected multiplex networks} 

\author{Kexian Zheng\inst{1 } \and Ying Liu\inst{1}\thanks{E-mail: \email{shinningliu@163.com}} \and Yang Wang \inst{1}  \and Wei Wang\inst{2}\thanks{E-mail: \email{wwzqbx@hotmail.com}}}
\shortauthor{Kexian Zheng \etal}

\institute{
  \inst{1} School of Computer Science, Southwest Petroleum University, Chengdu {\rm 610500}, China\\
  \inst{2} Cybersecurity Research Institute, Sichuan University, Chengdu {\rm 610065}, China\\
}

\pacs{89.75.Fb}{Structures and organization in complex systems}
\pacs{87.23.Ge}{Dynamics of social systems}
\pacs{89.75.-k}{Complex systems}

\abstract{
Many real-world networks are coupled together to maintain their normal functions. Here we study the robustness of multiplex networks with interdependent and interconnected links under \textbf{k}-core percolation, where a node fails when it connects to a threshold of less than \textbf{k} neighbors. By deriving the self-consistency equations, we solve the key quantities of interests such as the critical threshold and size of the giant component analytically and validate the theoretical results with numerical simulations. We find a rich phase transition phenomenon as we tune the inter-layer coupling strength. Specifically speaking, in the ER-ER multiplex networks, with the increase of coupling strength, the size of the giant component in each layer first undergoes a first-order transition and then a second-order transition and finally a first-order transition. This is due to the nature of inter-layer links with both connectivity and dependency simultaneously. The system is more robust if the dependency on the initial robust network is strong and more vulnerable if the dependency on the initial attacked network is strong. These effects are even amplified in the cascading process. When applying our model to the SF-SF multiplex networks, the type of transition changes. The system undergoes a first-order phase transition first only when the two layers' mutually coupling is very strong and a second-order transition in other conditions.}

\begin{document}
\maketitle
\section{\color{black}Introduction}
Most real-world networks are not isolated but are coupled with other systems to implement their functions well, which can be described by the multilayer network~\cite{bocalleti2014, mikko2014, domenico2013, gomez2013, gao2012,gao2014}. A typical example is the transportation network composed of airlines and railways, where different transportation ways support each other to maintain the traffic flow of people~\cite{barthelemy2011,halu2014}. Another example is the interdependent communication and power grid network, where a failure of a tiny fraction of nodes in one network induces the block down of the whole system~\cite{buldyrev2010}. The multilayer network approach has proved to be very successful in many fields of network science, such as resilience and robustness~\cite{gao2011}, spreading dynamics~\cite{wang2014,cozzo2013,domenico2016,arruda2017,jalili2017, wang2017}, synchronization~\cite{genio2016,zhuang2020}.

Real complex systems are frequently under random or intentional attacks, such as natural disasters of hurricanes, earthquakes, power outages, Internet router failure, and terrorist attacks~\cite{huang2011,gao2016}. These damages may crucially change or even destroy the structure and function of the network. Understanding the cascading induced by an initial failure is a critical question in the study of complex systems. Researches on network robustness is an active topic in network science~\cite{albert2000,callaway2000,wang2008,gao2012,yang2017}, where the robustness quantifies how resilient the network is under perturbations~\cite{schneider2011}.
In studying the robustness of networks, the ordinary percolation model is usually used~\cite{callaway2000}. In the ordinary percolation approach, initially, a fraction of $1-p$ of nodes are removed, which may disconnect more nodes from the largest connected components. If $1-p$ is large enough, then at a critical value $p_c$, the whole network collapses, and there are only negligible small clusters and isolated nodes. The size of the largest connected component in the remaining network is served as the order parameter that measures the robustness of the network structure and is used for studying the phase transition behaviors~\cite{feng2015}. A natural generalization of the ordinary percolation is the $k$-core percolation~\cite{doro2006}. The $k$-core is a highly interconnected part of the network, with each node in the $k$-core has at least $k$ neighbors~\cite{seidman1983}. It is obtained by removing the nodes with less than $k$ neighbors. If there appear new nodes with degree less than $k$, keep removing them until no further removal is possible. The $k$-core percolation implies the emergence of a giant $k$-core at a threshold of a proportion of nodes removed at random, which displays an entirely different phase transition from the ordinary percolation~\cite{goltsev2006}. These methods are generalized to study the robustness of multilayer networks~\cite{buldyrev2010,azimi2014,radicchi2015}.

Studies on the robustness of multilayer networks focus on two types of networks distinguished by their inter-layer coupling nature, which are the interdependent networks and the interconnected networks. In the interdependent networks, the presence of a node in one layer depends on a node's presence in the other layers. The removal of nodes on one layer leads to isolated nodes or small clusters in the same layer and leads to the removal of their dependent nodes in the other layers. Extensive studies have found that interdependency among layers makes the system more vulnerable~\cite{buldyrev2010, parshani2010, gao2011_2,baxter2012, yuan2016,panduranga2017,liu2019}. While in the interconnected networks, the connectivity links coupling the two layers provide an additional connection for the nodes, making the system significantly more robust~\cite{leicht2009,domenico2014, gross2020}. Besides, scientists studied the coupled networks with both interdependent and interconnected links, where the two kinds of links have competing effects of either decreasing or increasing the robustness of the system~\cite{hu2011, cao2020}. Understanding how the interdependency and interconnectivity impact the system's robustness is the main challenge in designing resilient infrastructures.

In some real-world scenarios, the coupling of networks may be interdependent as well as interconnected simultaneously. For example, in the social networks, an individual may have connections in different social media corresponding to other layers. To be active in the network, the actor should have a certain amount of effective connections within a layer or have connections in different layers. The individual's failure in one network leads to the failure of the individual in the other network because it is the same actor. In this case, the coupling of different social networks is interdependent and interconnected simultaneously.
Another example is the coupled financial systems, where individual banks are considered nodes. Their economic interplays such as credit, derivatives, foreign exchange, and securities are represented by links and grouped into layers by different types~\cite{seba2015}. A bank's business on one layer may support this bank's business in another layer due to its reputation in the first business. Still, the failure in one business may lead to the bankruptcy of the financial organization, and thus its activities in all layers are disabled.

In this paper, we study the robustness of coupled networks in which the inter-layer links have a nature of both dependency and connectivity simultaneously. Using the $k$-core percolation approach, we analyze the phase transition under random attacks, which agrees well with the simulation results. We find that in the ER-ER multiplex networks (i.e., each subnetwork is an  Erd\"{o}s-R\'{e}nyi (ER) network), with the increase of the asymmetrical coupling strength $q_a$ and $q_b$ between two layers, the interdependent and interconnected network first undergoes a first-order phase transition, and then a seconde order transition and finally a first-order transition again. Meanwhile, as the inter-layer links have a competing effect of either increasing the robustness due to the interconnectivity nature or decreasing the robustness due to the interdependency nature, we find that in a relatively large parameter ranges, the coupled network are more vulnerable when the coupling strength becomes strong, which means the dependency of links dominates in the cascading process. Specifically speaking, when the coupling strength $q_a$ is fixed, the robustness decreases with $q_b$. While $q_b$ is fixed, there are two stages. When $q_b$ is within $0.6$, the robustness increase with $q_a$, which means the connectivity of inter-layer links dominates in the cascading process. If $q_b$ is above $0.6$, the robustness is in general decrease with $q_a$. Finally, we study the coupled SF-SF multiplex networks (i.e., each subnetwork is scale-free (SF) network) and find that the phase transition phenomena are different from those of the coupled ER-ER networks.

The rest of the paper is organized as follows. In the model section, we describe the cascading model in the interdependent and interconnected networks. In the theory analysis section, we analyze the \textbf{k}-core percolation process in the coupled networks and derive the percolation threshold. In the simulation results section we demonstrate the simulation results and analyze the phase transition in the coupled ER-ER networks and SF-SF networks. A conclusion is given finally.

\section{The model}\label{sec:model}
Consider a system composed of two uncorrelated random networks A and B with the same number of nodes $N$ with distribution $P(k_a)$ and $P(k_b)$ respectively. A fraction $q_a$ of nodes in layer A depend on and are connected by nodes in layer B, which means for a node $i$ in layer A when its dependent node in layer B is functional, it provides one connection (one degree) to node $i$, but if the dependent node in layer B fails, node $i$ in layer A also fails. Thus the inter-layer link can be considered as directional with both dependency and connectivity. Similarly, a fraction $q_b$ of nodes in layer B depend on and are connected by nodes in layer A. We assume that each node in a layer depends on and is connected by at most one node in the other layer. $q_a$ and $q_b$ determines the coupling strength.
\begin{figure}[!t]
	\centering
	\onefigure[scale=.26]{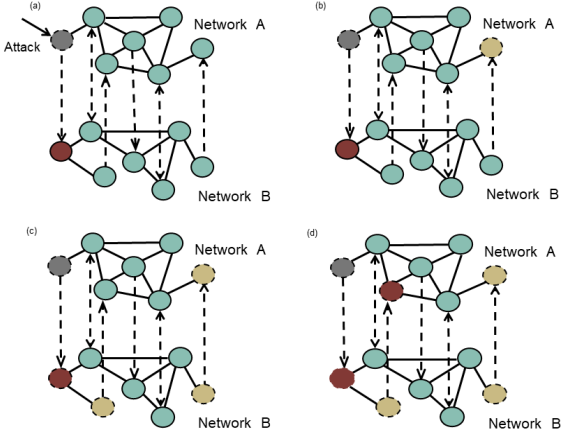}
	\caption{(Color online) Schematic representation of the cascading model. The dashed lines represent the inter-layer links directed from a node in one network to a node that depends on it. (a) Initially, the grey node is attacked, and the red node that is dependent on it is also removed. (b) $\mathbf{k}$-core pruning in network A, where the yellow node with neighbors few than $\phi_a$ is removed. If there are nodes in network B dependent on the yellow nodes, also remove them. (c) $\mathbf{k}$-core pruning in network B, where the yellow node with neighbors few than $\phi_b$ are removed. (d) The red node in network A that is dependent on the yellow nodes in network B is removed. Here $\phi_a=3$, $\phi_b=3$.}
	\label{figure1}
\end{figure}

Initially, a fraction $1-p$ of randomly chosen nodes in network A are removed, along with their edges. The nodes that are dependent on the removed nodes in layer B are also removed. In the classical $k$-core percolation, nodes in layer A with fewer neighbors than $\phi_a$, the local threshold, are removed, along with all the nodes in layer B dependent on them. Then in layer B, nodes with fewer neighbors than $\phi_b$ are removed and the nodes in layer A are dependent on them. This cascade process continues until a steady-state is reached. The system is either completely fragmented or a mutually connected giant \textbf{k}-core appears, where $\textbf{k}=(\phi_a, \phi_b)$~\cite{azimi2014}. Note that in the process of obtaining the \textbf{k}-core, a node $i$ in layer A that is dependent and connected by a node in layer B needs only ($\phi_a-1$) neighbors to be in the \textbf{k}-core, because the inter-layer link provides a connection to $i$. If a node in layer A has no inter-layer link, then it must have at least $\phi_a$ neighbors within layer A to be in the \textbf{k}-core. It is similar for nodes in layer B. The cascading process is demonstrated in Fig.~\ref{figure1}.

\section{Theoretical Analysis} \label{sec:theory}
In this section, we study the size of the giant \textbf{k}-core connected cluster and phase transition in the final sate. When the system reaches the final state in the cascading process, let $f_a$ be the probability that a given end of a randomly selected edge is the root of an infinite ($\phi_a-1$)-ary subtree. An end of an edge is a root of an infinite ($\phi_a-1$)-ary subtree if at least $\phi_a-1$ of its children are the roots of infinite ($\phi_a-1$)-ary subtrees, meanwhile the node it depends on must be in the $\phi_b$-core of network B. Similarly, $f_b$ is the probability that a given end of a randomly selected edge is the root of an infinite ($\phi_b-1$)-ary subtree. Then $f_a$ can be expressed as

\begin{footnotesize}
	\begin{equation}\label{fa}
	\begin{aligned}
	f_a&=\!p[(1\!-\!q_a)\!\sum_{k_a=\phi_a\!-\!1}^\infty \!\frac{P(k_a\!+\!1)(k_a\!+\!1)}{\langle k_a \rangle}\!\sum_{j=\phi_a\!-\!1}^{k_a}\!C_{k_a}^jf_a^{j}(1\!-\!f_a)^{k_a\!-\!j}\\
	&+q_a\!\sum_{k_a=\phi_a\!-\!2}^\infty\!\frac{P(k_a\!+\!1)(k_a\!+\!1)}{\langle k_a \rangle}\!\sum_{j=\phi_a\!-\!2}^{k_a}\!C_{k_a}^jf_a^j(1\!-\!f_a)^{k_a\!-\!j}\psi_b(f_b)],
	\end{aligned}
	\end{equation}
\end{footnotesize}

where $p$ is the probability that a node is not removed initially, and $C_{k_a}^j=k_a!/(k_a-j)!j!$. The term $P(k_a+1)(k_a+1)/\langle k_a \rangle $ is the probability that an end of an randomly chosen edge has $k_a$ out-going edges except the chosen edge, and $C_{k_a}^jf_a^{j}(1-f_a)^{k_a-j}$ is the probability that if a given end of an edge has $k_a$ children, then exactly $j$ of them are the roots of ($\phi_a-1$)-ary subtrees~\cite{doro2006}. In Eq.~(\ref{fa}), two types of nodes are represented. The first type is the nodes with no inter-layer links, corresponding to the term with coefficient $(1-q_a)$, where these nodes should have no less than $\phi_a$ neighbors within network A. The second type is the nodes with inter-layer links, corresponding to the term with coefficient $q_a$, where these nodes should have no less than $\phi_a-1$ neighbors within network A, and the remaining one neighbors is the node it is dependent on and connected to in the $\phi_b$-core of network B. The term $\psi_b(f_b)$ is the probability of a node in the $\phi_b$-core of network B in the steady state, where it is expressed as

\begin{footnotesize}
\begin{equation}\label{fiBfB}
\begin{aligned}
\psi_b(f_b)&=(1-q_b)\sum_{k_b=\phi_b}^\infty P(k_b)\sum_{j=\phi_b}^{k_b}C_{k_b}^jf_b^j(1-f_b)^{k_b-j}\\
&+q_b\sum_{k_b=\phi_b-1}^\infty P(k_b)\sum_{k_b=\phi_b-1}^{k_b}C_{k_b}^jf_b^j(1-f_b)^{k_b-j}.
\end{aligned}
\end{equation}
\end{footnotesize}

In Eq.~(\ref{fiBfB}), the two terms on the right respectively represent the probability of nodes without or with inter-layer links in the $\phi_b$-core in network B.

Similarly, we can obtain the equation of $f_b$ as

\begin{footnotesize}
\begin{equation}\label{fb}
\begin{aligned}
f_b&=\!(1\!-\!q_b)\!\sum_{k_b=\phi_b\!-\!1}^\infty \!\frac{P(k_b\!+\!1)(k_b\!+\!1)}{\langle k_b \rangle}\!\sum_{j=\phi_b\!-\!1}^{k_b}C_{k_b}^jf_b^{j}(1\!-\!f_b)^{k_b\!-\!j}\\
&+pq_b\!\sum_{k_b=\phi_b\!-\!2}^\infty\!\frac{P(k_b\!+\!1)(k_b\!+\!1)}{\langle k_b \rangle}\!\sum_{j=\phi_b\!-\!2}^{k_b}C_{k_b}^jf_b^j(1\!-\!f_b)^{k_b\!-\!j}\psi_a(f_a)
\end{aligned}
\end{equation}
\end{footnotesize}

In the second term on the right of the equation, the multiplier $p$ represents the probability that an end of the edge is occupied (not removed initially). $\psi_a(f_a)$ is the probability that a randomly chosen node belongs to the $\phi_a$-core in network A, which is
\begin{footnotesize}
\begin{equation}\label{fiAfA}
\begin{aligned}
\psi_a(f_a)&=(1-q_a)\sum_{k_a=\phi_a}^\infty P(k_a)\sum_{j=\phi_a}^{k_a}C_{k_a}^jf_a^j(1-f_a)^{k_a-j}\\
&+q_a\sum_{k_a=\phi_a-1}^\infty P(k_a)\sum_{k_a=\phi_a-1}^{k_a}C_{k_a}^jf_a^j(1-f_a)^{k_a-j}.
\end{aligned}
\end{equation}
\end{footnotesize}

For any given value of $p$, the $f_a$ and $f_b$ can be solved from Eqs.~(\ref{fa}) and (\ref{fb}) using the Newton's method~\cite{yuan2016} after giving appropriate initial values.

We denote $M_{\phi_a}^A(p)$ and $M_{\phi_b}^B(p)$ as the probability of a randomly chosen node in network A or B belongs to the mutually connected ($\phi_a$, $\phi_b$)-core, which satisfy

\begin{footnotesize}
\begin{equation}\label{ma}
\begin{aligned}
M_{\phi_a}^A(p)&=p[(1-q_a)\sum_{k_a=\phi_a}^\infty P(k_a)\sum_{j=\phi_a}^{k_a}C_{k_a}^jf_a^{j}(1-f_a)^{k_a-j}\\
&+q_a\sum_{k_a=\phi_a-1}^\infty P(k_a)\sum_{j=\phi_a-1}^{k_a}C_{k_a}^jf_a^j(1-f_a)^{k_a-j}\psi_b(f_b)],
\end{aligned}
\end{equation}
\end{footnotesize}
and
\begin{footnotesize}
\begin{equation}\label{mb}
\begin{aligned}
M_{\phi_b}^B(p)&=(1-q_b)\sum_{k_b=\phi_b}^\infty P(k_b)\sum_{j=\phi_b}^{k_b}C_{k_b}^jf_b^{j}(1-f_b)^{k_b-j}\\
&+pq_b\sum_{k_b=\phi_b-1}^\infty P(k_b)\sum_{j=\phi_b-1}^{k_b}C_{k_b}^jf_b^j(1-f_b)^{k_b-j}\psi_a(f_a).
\end{aligned}
\end{equation}
\end{footnotesize}

\begin{figure}[!t]
	\centering
	\onefigure[scale=.22]{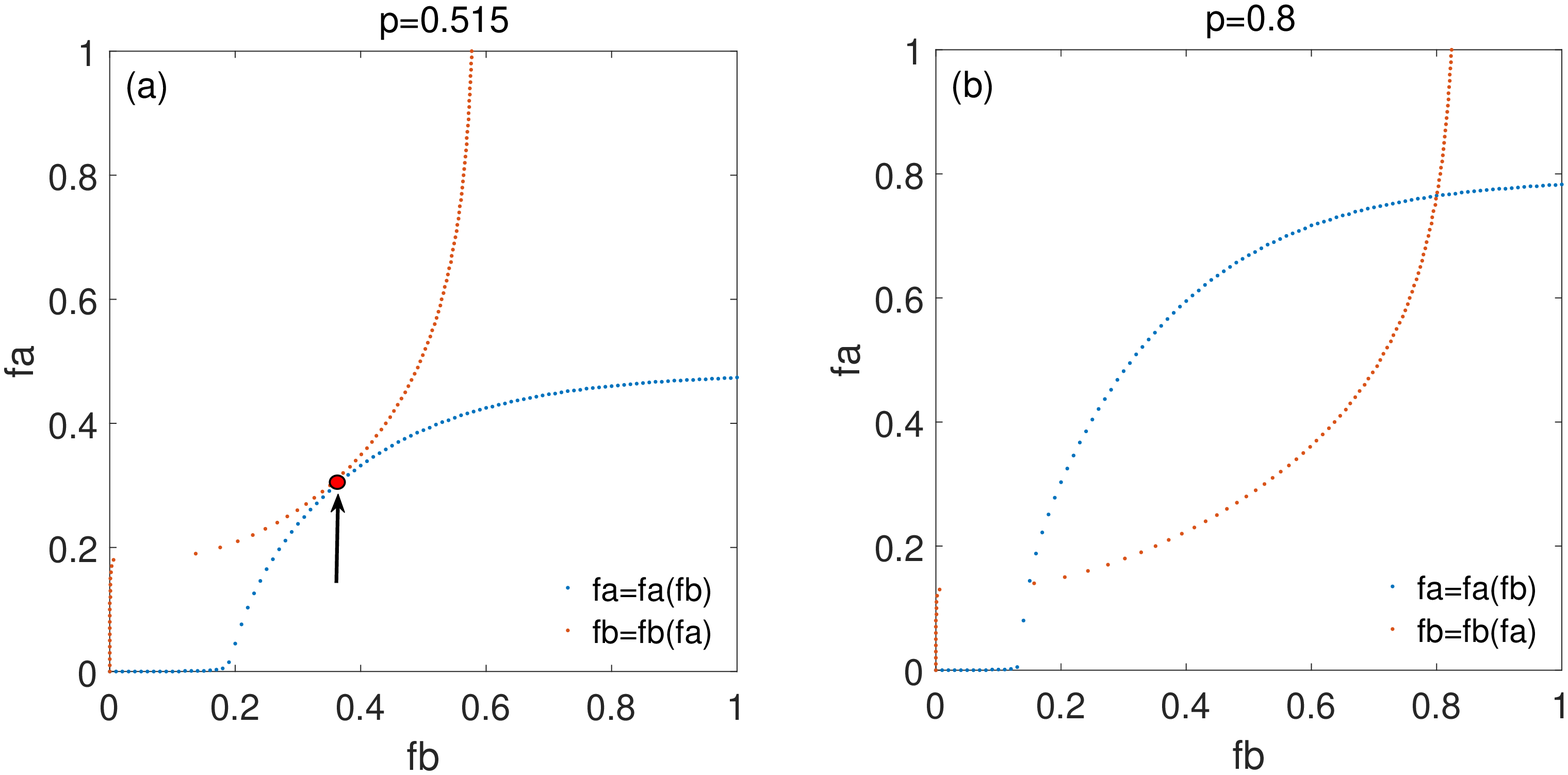}
	\caption{Graphical solution of Eqs.~(\ref{fa}) and (\ref{fb}). (a) At $p=p_c=0.515$, the curves of $f_a(f_b)$ and $f_b(f_a)$ tangentially touch each other at the critical point. (b) When $p>p_c$, the two curves have two intersections and the larger one corresponds to the solution of Eqs.~(\ref{fa}) and (\ref{fb}) where the giant \textbf{k}-core exists. The other parameters are set as $\phi_a=\phi_b=3$, $q_a=q_b=0.8$, and the mean degree $\langle k_a \rangle=\langle k_b \rangle=7$.}
	\label{figure2}
\end{figure}

The solutions of Eqs.~(\ref{fa}) and (\ref{fb}) can be graphically represented on a $f_a$, $f_b$ plane~\cite{parshani2010}. For small values of $p$, the Eqs.~(\ref{fa}) and (\ref{fb}) has the trivial solution of $f_a=f_b=0$, which implies the absence of \textbf{k}-core in the system. As $p$ increases, at a critical value $p=p_c$, the mutually connected giant \textbf{k}-core appears. In this case, the two curves $f_a=f_a(f_b)$ and $f_b=f_b(f_a)$ tangentially touch each other at a point, and meet the condition

\begin{equation}\label{tan}
\begin{aligned}
\frac{df_a}{df_b}\cdot \frac{df_b}{df_a}=1,
\end{aligned}
\end{equation}

which implies a first-order transition at the touching point as shown in Fig.~\ref{figure2} (a). When $p>p_c$, shown in Fig.~\ref{figure2} (b), the two curves will always have non-zero intersections. The larger value of $f_a$ and $f_b$ is the physical solutions of Eqs.~(\ref{fa}) and (\ref{fb}) where the giant \textbf{k}-core exists. For the smaller one, as it is under the value at the the threshold, there is no giant \textbf{k}-core and this solution is physically meaningless.

\section{Simulation results}\label{sec:simulations}
\subsection{ ER-ER multiplex networks}
\begin{figure}[!t]
	\centering
	\onefigure[scale=.36]{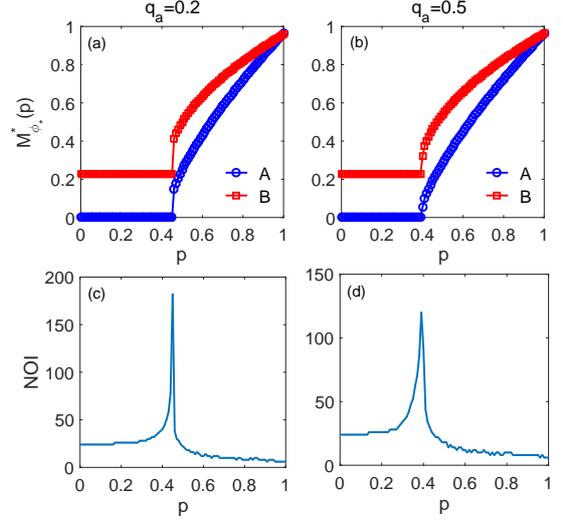}
	\caption{Size of the \textbf{k}-core of network A $M_{\phi_a}^A(p)$ and network B $M_{\phi_b}^B(p)$ and the number of iterations in obtaining $f_a$ and $f_b$ as a function of $p$. (a) Size of $k$-core as a function of p, where $q_a=0.2$ and $q_b=0.5$. (b) Size of \textbf{k}-core as a function of $p$, where $q_a=0.5$ and $q_b=0.5$. (c) NOI as a function of $p$, where $q_a=0.2$ and $q_b=0.5$. (d) NOI as a function of $p$, where $q_a=0.5$ and $q_b=0.5$. The solid lines are theoretical predictions and the symbols are simulation results, which agree very well.}
	\label{figure3}
\end{figure}

We construct the interdependent and interconnected ER-ER multiplex networks, where the network size is $N_a=N_b=10000$ and the degree distribution of each network is $P(k)= e^{-\lambda}\frac{\lambda ^k}{k!}$, and $\lambda_a=\lambda_b=7$ is the average degree. We set $\phi_a=\phi_b=3$ as the local threshold to obtain \textbf{k}-core. The \textbf{k}-core percolation process starts by randomly removing a fraction $1-p$ of nodes in network A, and the pruning process continues until the steady state is reached. Fig.~\ref{figure3} shows the giant connected cluster size of the \textbf{k}-core of network A and network B, and the number of iterations (NOI) in obtaining $f_a$ and $f_b$ as a function of $p$. It can be seen from Figs.\ref{figure3} (a) and (b) that when the coupling strength is weak, for $q_a=0.2$, both $M_{\phi_a}^A(p)$  and $M_{\phi_b}^B(p)$  undergo a discontinuous first-order transition. As the coupling strength becomes strong, for $q_a=0.5$, the transition becomes a continuous second-order transition. Increasing the number of inter-layer links makes the system more robust and more controllable. The number of iterations in obtaining $f_a$ and $f_b$ from Eqs.~(\ref{fa}) and (\ref{fb}) reaches a maximum at the percolation threshold $p_c$, as shown in Figs.~\ref{figure3} (c) and (d).

\begin{figure}[!t]
	\centering
	\onefigure[scale=.29]{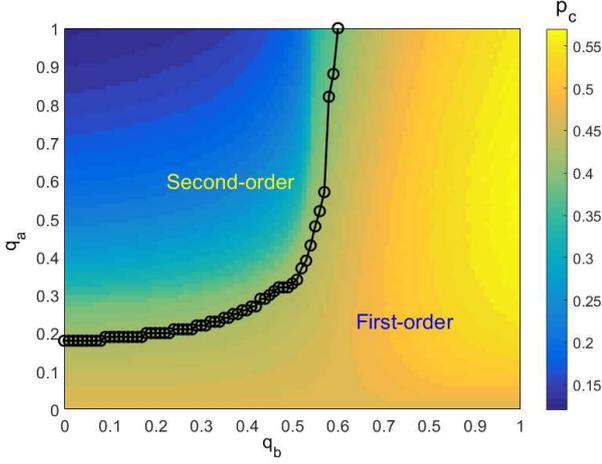}
	\caption{Phase diagram of $p_c$ as a function of $q_a$ and $q_b$. The black line separates the region of first-order transition and second-order transition.}
	\label{figure4}
\end{figure}

We demonstrate the phase diagram of the percolation in network A as a function of the coupling strength $q_a$ and $q_b$ in Fig.~\ref{figure4}.
The two types of phase transition is separated by the black lines. To determine the boundary of the first-order and second-order transition, we use the following method~\cite{cao2020}. Consider $p$ increases from $0$ to $1$. At the percolation threshold $p=p_c$, there is a first-order or second-order transition. Define $F_a(p, f_a, f_b, \psi_b)$ as the right side of Eq.~(\ref{fa}) and $F_b(p, f_a, f_b, \psi_a)$ as the right side of Eq.~(\ref{fb}). When $p$ is approaching the second-order transition threshold $p_c^{\textrm{II}}$, $f_a,f_b,\psi_a,\psi_b \rightarrow 0$. Do the Taylor expansion of $F_a$, we obtain
\begin{equation}\label{FaTaylor}
\begin{aligned}
fa=F_a^{'}(p_c^{\textrm{II}},0,0,0)fa+\frac{1}{2}F_a^{''}(p_c^{\textrm{II}},0,0,0)f_a^2+O(f_a^3).
\end{aligned}
\end{equation}
Dividing $f_a$ on both sides of Eq.~(\ref{FaTaylor}) comes to
\begin{equation}\label{FaTaylor_div}
\begin{aligned}
1=F_a^{'}(p_c^{\textrm{II}},0,0,0)+\frac{1}{2}F_a^{''}(p_c^{\textrm{II}},0,0,0)f_a+O(f_a^3).
\end{aligned}
\end{equation}
As $f_a\rightarrow 0$, we neglect the second and third terms on the right side of Eq.~\ref{FaTaylor_div} and get $1=F_a^{'}(p_c^{\textrm{II}},0,0,0)$, where the threshold at the second-order transition $p_c^{\textrm{II}}$ can be calculated. As for the case when $p$ is approaching the first-order transition threshold $p_c^{\rm I}$, there is a jump for $f_a, f_b, \psi_a, \psi_b$ from $0$ to non-zero values. The non-trivial solutions $f_{a_c}$ and $f_{b_c}$ at this point satisfies
\begin{equation}\label{Fa_partial}
\begin{aligned}
\frac{\partial F_a(p, f_a, f_b, \psi_b)}{\partial f_a}|_{f_a=f_{a_c}, f_b=f_{b_c}, p=p_c^{\rm I}}=1.
\end{aligned}
\end{equation}
and
\begin{equation}\label{Fb_partial}
\begin{aligned}
\frac{\partial F_b(p, f_a, f_b, \psi_b)}{\partial f_b}|_{f_a=f_{a_c}, f_b=f_{b_c}, p=p_c^{\rm I}}=1.
\end{aligned}
\end{equation}
Combing Eqs.~(\ref{fa})-(\ref{fiAfA}) and (\ref{Fa_partial})-(\ref{Fb_partial}), the threshold for first-order transition $p_c^{\rm I}$ can be solved.
At the boundary, $p_c^{\rm I}=p_c^{\textrm{II}}=p_c$, where $p_c$ satisfies both conditions for determining the first-order threshold and second-order threshold. Take $1=F_a^{'}(p_c,0,0,0)$ into Eq.~(\ref{FaTaylor}), we obtain

\begin{equation}\label{FaTaylor_2}
\begin{aligned}
\frac{1}{2}F_a^{''}(p_c,0,0,0)f_a^2+O(f_a^3)=0.
\end{aligned}
\end{equation}

As $f_a$ and $f_b$ has non-trivial solution at the first-order transition threshold, then

\begin{equation}\label{FaTaylor_3}
\begin{aligned}
F_a^{''}(p_c,0,0,0)=0.
\end{aligned}
\end{equation}

By solving Eq.~\ref{FaTaylor_3}, the boundary of the first-order threshold and second-order threshold can be obtained.

From Fig.~\ref{figure4} it can be seen that when $q_a<0.2$, $p_c$ increases with $q_b$ and there is only first-order transition. When $q_a>0.2$, $p_c$ also increases with $q_b$, and the phase transition is a second-order then followed by a first-order transition. As for $q_b$, when it is smaller than $0.6$, with the increase of $q_a$, the $p_c$ decreases, which reflects an increased system robustness. When $q_b>0.6$, with the increase of $q_a$, $p_c$ increases first and then decrease a little bit. The above phenomena can be explained as follows. For a fixed $q_a$, the increase of $q_b$ means stronger dependency and connectivity of network B on network A. As network A is initially under random attack, the larger dependency, the more nodes in network B will be removed at the first step. This will cause larger removed nodes in network A and B in the following cascading process, thus making the system more vulnerable. While for a fixed $q_b$ not too large, with the increase of $q_a$, the more nodes in network A are dependent on and connected by nodes from network B. As initially all nodes in network B are functional, the increasing dependency and connectivity on network B makes more nodes in network A receive one more connection from network B thus making network A more robust, which also increase the robustness of the system. We can consider this simply as increasing dependency on the initially attacked network A makes the system more vulnerable, while increasing dependency on the network B makes system more robust.

\begin{figure}[!t]
	\centering
	\onefigure[scale=.36]{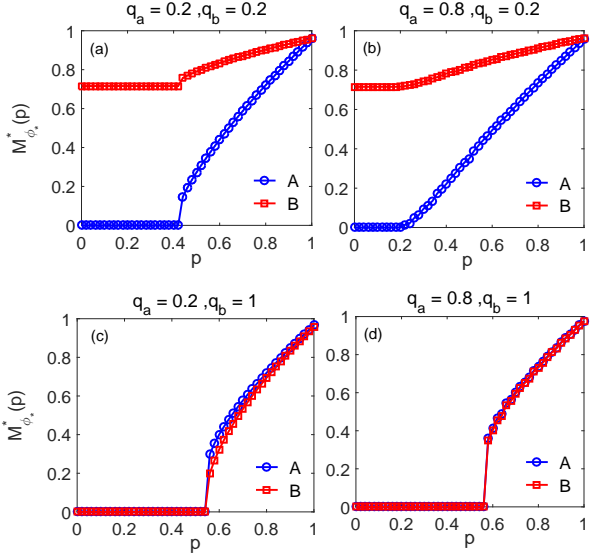}
	\caption{Size of the giant connected component of \textbf{k}-core in network A and network B as a function of $p$. (a) Network A undergoes a first-order transition at $q_a=0.2$, $q_b=0.2$, $p_c=0.42$. (b) Network A undergoes a second-order transition at $q_a=0.8$, $q_b=0.2$, $p_c=0.16$ (c) Both networks undergo a first-order transition at $q_a=0.2$, $q_b=1.0$, $p_c=0.54$. (d) Both networks undergo a first-order transition at $q_a=0.8$, $q_b=1.0$, $p_c=0.57$. The solid lines are theoretical predictions and the symbols are simulation results.}
	\label{figure5}
\end{figure}

We compare the phase transitions under different combinations of coupling strength $q_a$ and $q_b$, as shown in Fig.~\ref{figure5}. In Figs.~\ref{figure5} (a) and (b), it can be seen that increasing the coupling strength $q_a$ reduces the percolation threshold $p_c$ and makes the network more robust when $q_b$ is relatively small. This is because network B is weakly impacted by network A, and more nodes in network A are connected by inter-layer links from B with the increase of $q_a$. In this case the network B helps to maintain the robustness of network A. But if network B is impacted strongly by network A, corresponding to the large $q_b$ as shown in (c) and (d), the damaged network B further reduces the robustness of network A as $q_a$ increases. If we compare Figs.~\ref{figure5} (a) and (c), as well as (b) and (d), we can find that no matter how large $q_a$ is, the system becomes more vulnerable when the dependency of network B on network A $q_b$ increases. These results are consistent with our analysis on Fig.~\ref{figure4}.

\begin{figure}[!t]
	\centering
	\onefigure[scale=.218]{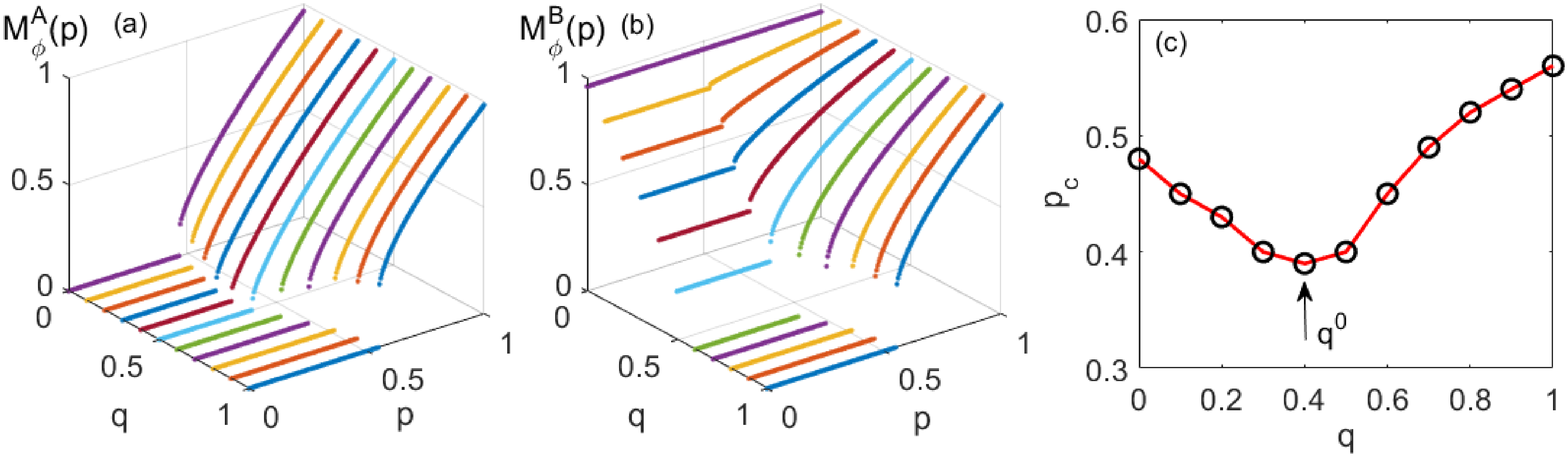}
	\caption{ Size of the giant connected component of \textbf{k}-core in network A (a) and network B (b) as a function of removed fraction $p$ and the symmetrical coupling strength $q$. In network A, the sequence of phase transition is first-order, second-order and first-order with the increase of $q$. While in network B, the phase transition is second-order and then first-order. (c) $p_c$ as a function of $q$. There is an optimal $q^o$ at which the network is the most robustness. The $p_c$ is obtained by simulations from Eqs.(1)-(6).}
	\label{figure6}
\end{figure}

Finally we focus on the case of symmetrical coupling of $q_a=q_b=q$ on the robustness of network A. From Fig.~\ref{figure6} (a), we can see that when $q$ increases from $0$ to $0.4$, the threshold $p_c$ decreases, and the network A undergoes a first-order transition first and then becomes a second-order transition. The increase of inter-layer links boost the robustness of the system. When $q$ increases from $0.4$ to $1.0$, the threshold $p_c$ begin to increase, and the phase transition changes to the first-order transition again. The increase of inter-layer links reduce the robustness of the system when the coupling becomes strong. As for network B, the increase of $q$ makes the transition changes from second-order to first-order, which implies the dependency on the initial attacked network makes the system more vulnerable. There is an optimal coupling strength $q^o$, at which $p_c$ is the smallest and the network A is the most robustness. This implies a mediate coupling strength best balance the effects of dependency and connectivity.

\subsection{SF-SF multiplex networks}
\begin{figure}[!t]
	\centering
	\onefigure[scale=.28]{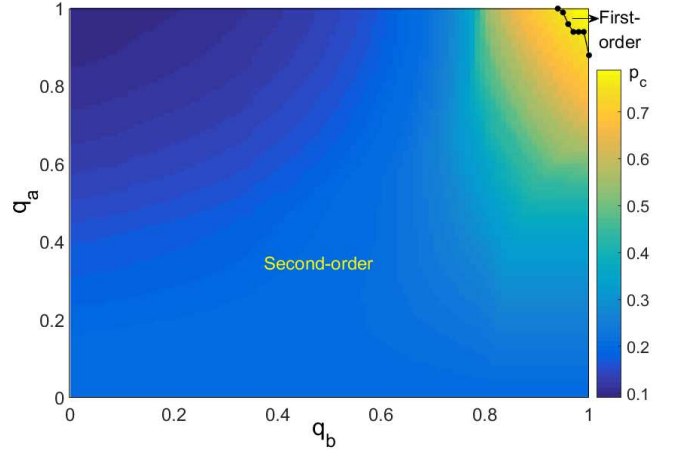}
	\caption{Phase diagram of $p_c$ as a function of $q_a$ and $q_b$ for network A. The black line separates the region of the first-order transition and second-order transition. In most of the diagrams, network A undergoes a first-order transition. When both $q_a$ and $q_b$ are above $0.95$, the network undergoes a first-order transition. $N_a=N_b=10000$, $k_{\rm min}=2$, $k_{\rm max}=50$, $\lambda=2.3$, $\phi_a=\phi_b=3$.}
	\label{figure7}
\end{figure}

Next we construct two networks that both follow the power law degree distribution $p(k)\sim k^{-\lambda}$, and connect them by inter-layer links, where $\lambda$ is the degree exponent. Fig.\ref{figure7} shows the phase diagram of k-core percolation transition of network A. It can be seen that the percolation under most of the coupling strength $q_a$ and $q_b$ is second-order, which is quite different from that of the ER networks. This is because in the ER networks, a large number of nodes have a degree around the mean degree thus are above the local threshold. When there are inter-layer links, the nodes are much impacted by the interdependency and interconnectivity. For the ER networks, the strong dependency on the initially robust network B, corresponding to large $q_a$ and small $q_b$, makes the system undergo a second-order transition, compared to that of the first-order transition in the single ER networks~\cite{yuan2016}. While for the SF networks, a large number of nodes have a small degree and will be pruned in the \textbf{k}-core percolation process. Thus the system is less impacted by the inter-layer coupling. The coupled network undergoes a second-order transition, which is similar to that of the single SF networks, corresponding to $q_a=q_b=0$. When the coupling is very strong, corresponding to large $q_a$ and $q_b$, the system undergoes a first-order transition.

\section{\color{black}Conclusion}\label{sec:discussion}
In this work we have studied the robustness of the coupled networks by \textbf{k}-core percolation, where the inter-layer links with the nature of interdependency and interconnectivity either increase or decrease the robustness of the system under different coupling strength. As the coupling is directed, the interdependency and interconnectivity have competing effects in impacting the robustness. We find that the strong dependency on the initial attacked network leads to a more vulnerable coupled system, while the strong dependency on the initial robust network leads to a more robust system. When increasing the mutual coupling strength of both layers, the system undergoes a first-order transition first, and then a second-order transition, followed by a first-order transition. There is an optimal coupling strength where the percolation threshold is the lowest, and the system is the most robust. When applied our model on the coupled SF networks, the phase transition is different from that of the coupled ER due to the local structural characteristics of the networks. As the SF networks are robust under random attack, using the SF networks as the coupled network better increases the robustness of an initially attacked ER network under some coupling strength.

\acknowledgments
This work is supported by the National Natural Science Foundation of China (Nos. 61802321 and 61903266),  Sichuan Science and Technology Program (Nos. 2020YJ0125 and 2020YJ0048), China Postdoctoral Science Special Foundation (No. 2019T120829), Fundamental Research Funds for the Central Universities, and the Southwest Petroleum University Innovation Base (No.642).

\end{document}